\documentclass[journal, twocolumn, final]{IEEEtran}

\usepackage[noadjust]{cite}

\ifCLASSINFOpdf
  \usepackage[pdftex]{graphicx}

\else
  \usepackage[dvips]{graphicx}
\fi

\usepackage{amsmath}
\usepackage{array}

\usepackage[caption=false,font=normalsize,labelfont=sf,textfont=sf]{subfig}

\usepackage{fixltx2e}

\usepackage{stfloats}

\ifCLASSOPTIONcaptionsoff
  \usepackage[nomarkers]{endfloat}
 \let\MYoriglatexcaption\caption
 \renewcommand{\caption}[2][\relax]{\MYoriglatexcaption[#2]{#2}}
\fi

 \let\MYorigsubfloat\subfloat
 \renewcommand{\subfloat}[2][\relax]{\MYorigsubfloat[]{#2}}

\begin{document}

\title{Switching Mechanism and the Scalability of vertical-TFETs}

\author{Fan Chen, Hesameddin Ilatikhameneh, Yaohua Tan, Gerhard Klimeck and Rajib Rahman

\thanks{The authors are with the Network for Computational Nanotechnology (NCN), Purdue University, West Lafayette, IN, 47906 USA. E-mail: fanchen@purdue.edu.}
\thanks{This work was supported in part by the Center for Low Energy Systems Technology, one of six centers of STARnet, and in part by the Semiconductor Research Corporation Program through Microelectronics Advanced Research Corporation and Defense Advanced Research Projects Agency. }}


\maketitle

\begin{abstract}
In this work, vertical tunnel field-effect transistors (v-TFETs) based on vertically stacked heretojunctions from 2D transition metal dichalcogenide (TMD) materials are studied by atomistic quantum transport simulations. The switching mechanism of v-TFET is found to be different from previous predictions. As a consequence of this switching mechanism, the extension region, where the materials are not stacked over is found to be critical for turning off the v-TFET. This extension region makes the scaling of v-TFETs challenging. In addition, due to the presence of both positive and negative charges inside the channel, v-TFETs also exhibit negative capacitance. As a result, v-TFETs have good energy-delay products and are one of the promising candidates for low power applications.
\end{abstract}


\IEEEpeerreviewmaketitle

\section{Introduction}

Tunneling Field Effect Transistors (TFETs) have been one the major device candidates for reducing the supply voltage $V_{DD}$ and power consumption since they can overcome the thermionic limitation of sub-threshold swing (SS) \cite{avci2015tunnel, appenzeller2004band, taur2015analytic}. The low ON current issue of TFETs has been shown to have several solutions such as III-V\cite{lu2012performance, zhou2012novel, fay2017iii}, 2D materials \cite{ roy20162d, ilatikhameneh2016saving, fiori2014electronics, chen2017thickness, liu2016atomistic, 7301966, liu2014ballistic, chen2016configurable, fiori2009ultralow, chen2015transport} and vertical tunnel field effect transistors (v-TFETs) based on tunneling through vertically stacked heterojunctions \cite{sarkar2015subthermionic, v17, v16, v15, v11, li2017characteristics, roy2015dual}. Although, a prototype v-TFET has been experimentally fabricated and shown to have a steep slope \cite{sarkar2015subthermionic}, it still suffers from low ON current. 2D materials can serve as ideal channel materials for v-TFETs due to their excellent device electrostatics \cite{v23}, lack of dangling bonds at interfaces and the variety choice of material selection. TFET designs exploiting different 2D material properties\cite{ ilatikhameneh2016saving, lam2010simulation, chen2017thickness, fiori2009ultralow, chen2016configurable} have been simulated to have high ON currents. Theoretical works also show a high ON current in v-TFETs by the choice of 2D material combination that forms staggered or broken band gap heterojunctions\cite{v16, v17}. However, a clear understanding of the switching mechanism, the important design aspects and scalability of v-TFETs subjected to the constraints of small SS and large ON-current are still missing. 

\begin{figure}[!t]
\centering
\subfloat[]{\includegraphics[width=0.22\textwidth]{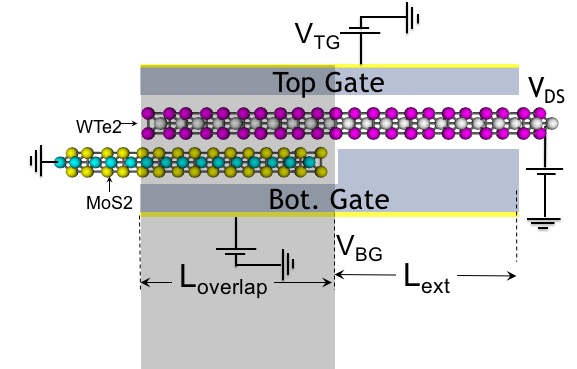}%
\label{1a}}
\hfil
\subfloat[]{\includegraphics[width=0.22\textwidth]{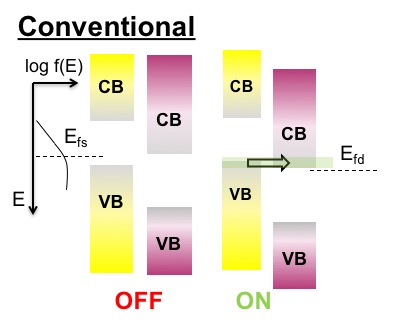}%
\label{1b}}
\vfil
\subfloat[]{\includegraphics[width=0.45\textwidth]{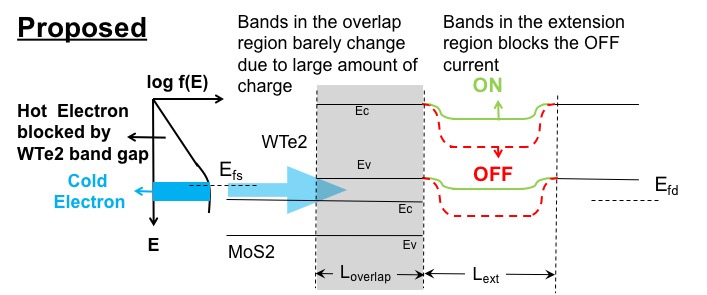}%
\label{1c}}

\caption{(a) The device structure of $MoS_2$-$WTe_2$ v-TFET. (b) The conventionally accepted switching mechanism of v-TFETs. (c) The switching mechanism proposed in this study.}
\label{1}

\end{figure}

  In this work, a vertical TFET based on vertically stacked $MoS_2$ and $WTe_2$ as shown in Fig.~\ref{1a} is studied. The switching mechanism of such v-TFETs has been widely accepted to be electrical detuning of relative band edges (or "band shifting") \cite{sarkar2015subthermionic, v17, v16, v15, v11, li2017characteristics}, where the device is turned on and off by inducing or removing a tunneling energy window between top and bottom layers respectively as shown in Fig.~\ref{1b}. In contrast, the atom-resolved charge distribution and band diagram obtained from quantum transport simulations disclose that the bands in the overlap region are hardly shifted by the gates.  Due to the screening from the excessive charge, the gate mainly impacts the potential at the extension region which is the key element for switching off the v-TFET. Hence, the switching in v-TFET is due to energy filtering in horizontal direction as shown in Fig.~\ref{1c}. A detailed discussion about this switching mechanism and Fig.~\ref{1c} can be found in section III. Simulations also show an extension region with $L_{ext}$ that is longer than 10nm is necessary to keep both the OFF current and SS small. As a result, v-TFETs cannot be scaled easily as it promised. Due to the  simultaneous presence of opposite charges inside the channel, the v-TFET is also found to have a negative capacitance. This compensates the degradation of ON current due to the scaling of the overlap region. \\
  
\section{Method}

The model assumptions and the band structure comparisons of tight-binding with DFT can be found in \cite{chen2017transport, tan2016first}. The transport characteristics of the $MoS_2$-$WTe_2$ v-TFET have been simulated using the self-consistent Poisson-semiclassical density approach to obtain the potential profile and then this potential is input to quantum transmitting boundary method (an equivalent of non-equilibrium Green$'$s function method for ballistic transport) in the multi-scale \cite{chen2015nemo5} and multi-physics \cite{chen2015surface, 24} Nano-Electronic MOdeling (NEMO5) tool  \cite{fonseca2013efficient}. A source and drain doping of $10^{20} cm^{-3}$ is used. Equivalent oxide thickness (EOT) is set to 0.5nm. 
\section{Results and Discussion}

\begin{figure}
\centering
\subfloat[]{\includegraphics[width=0.38\textwidth]{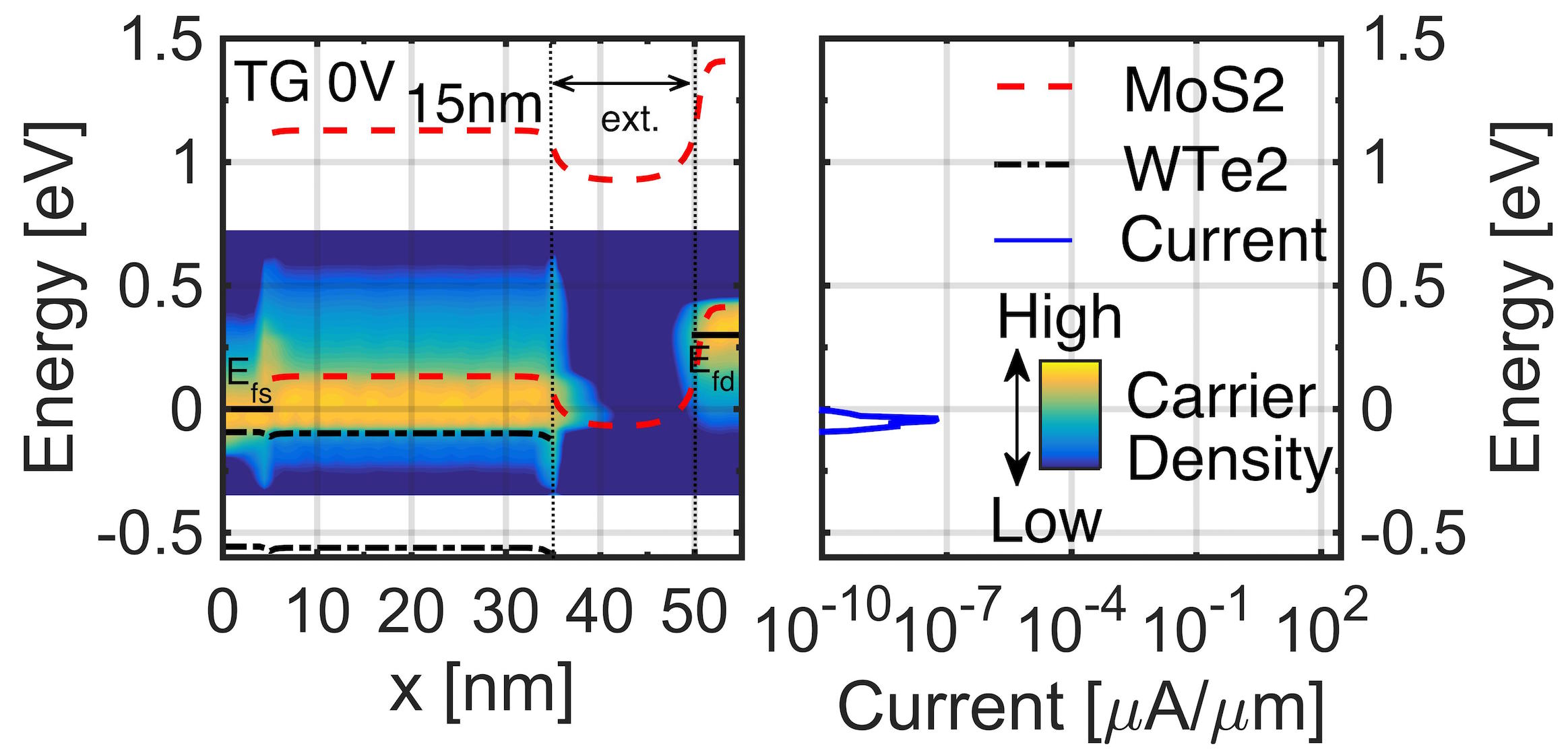}%
\label{2a}}
\vfil
\subfloat[]{\includegraphics[width=0.38\textwidth]{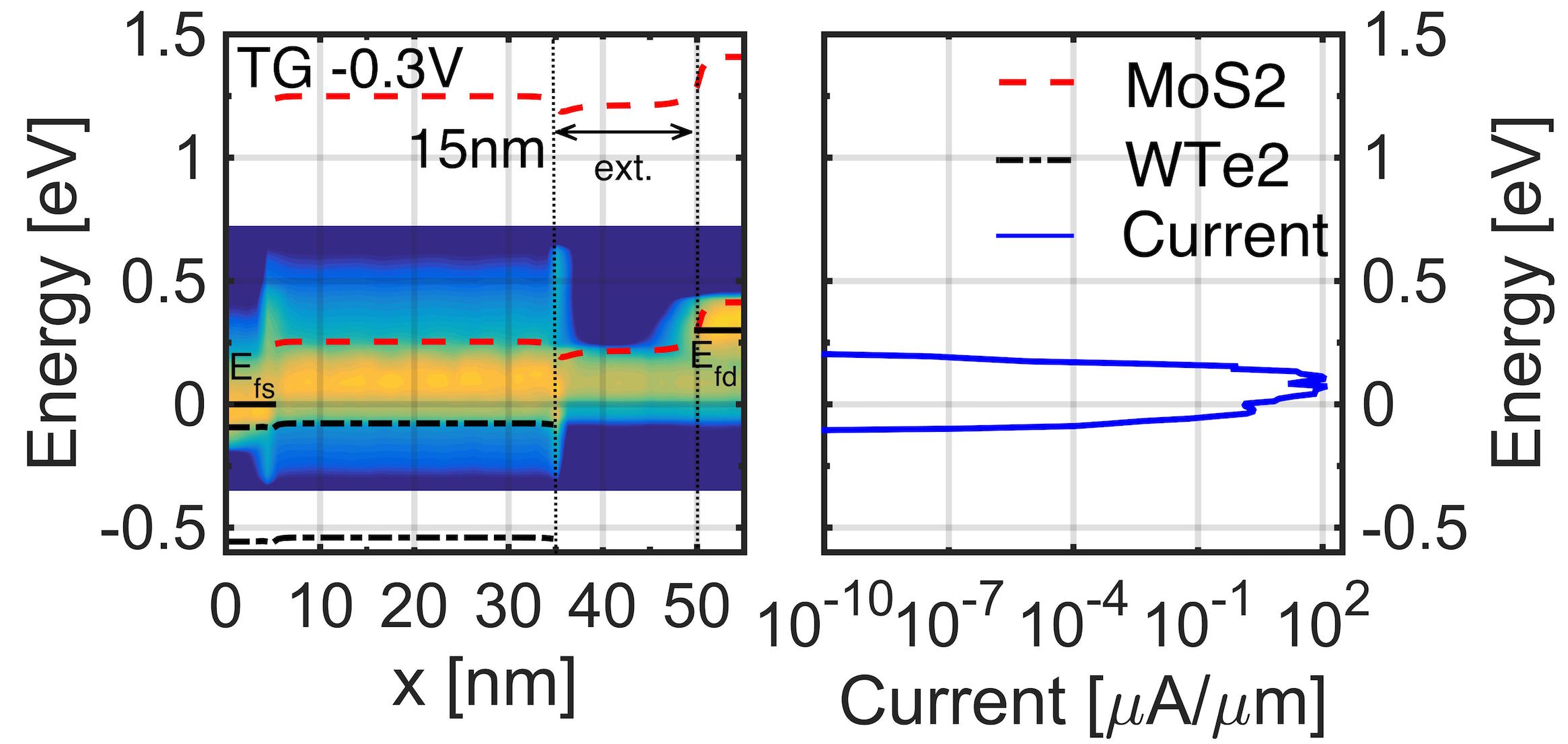}%
\label{2b}}
\caption{ \label{label} The band edges of a $MoS_2$-$WTe_2$ interlayer TFET at (a) OFF and (b) ON state aligned with the energy resolved current respectively.}
\label{2}
\end{figure}

\begin{figure}[!t]
\centering
\subfloat[]{\includegraphics[width=0.2\textwidth]{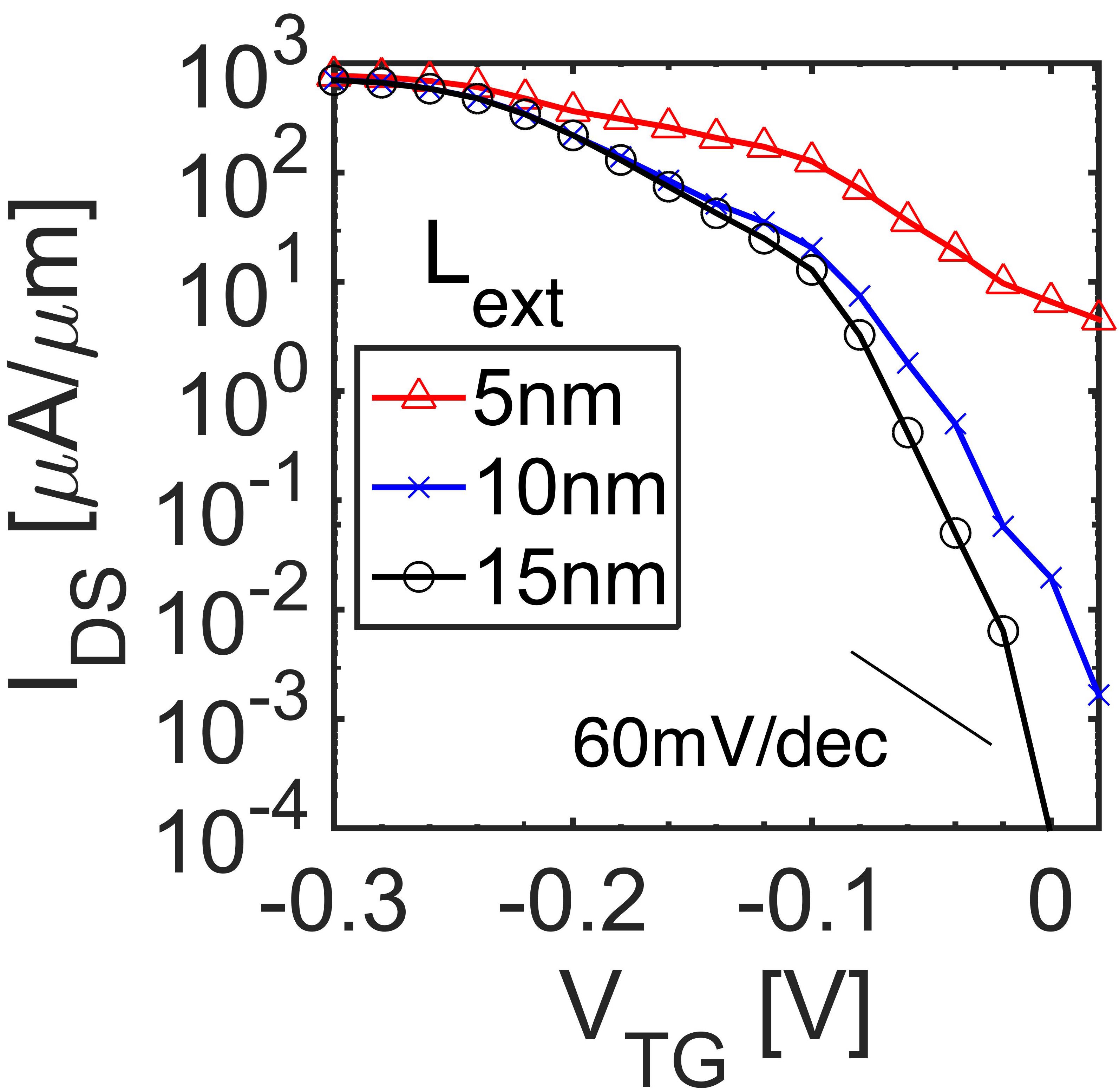}%
\label{3a}}
\hfil
\subfloat[]{\includegraphics[width=0.19\textwidth]{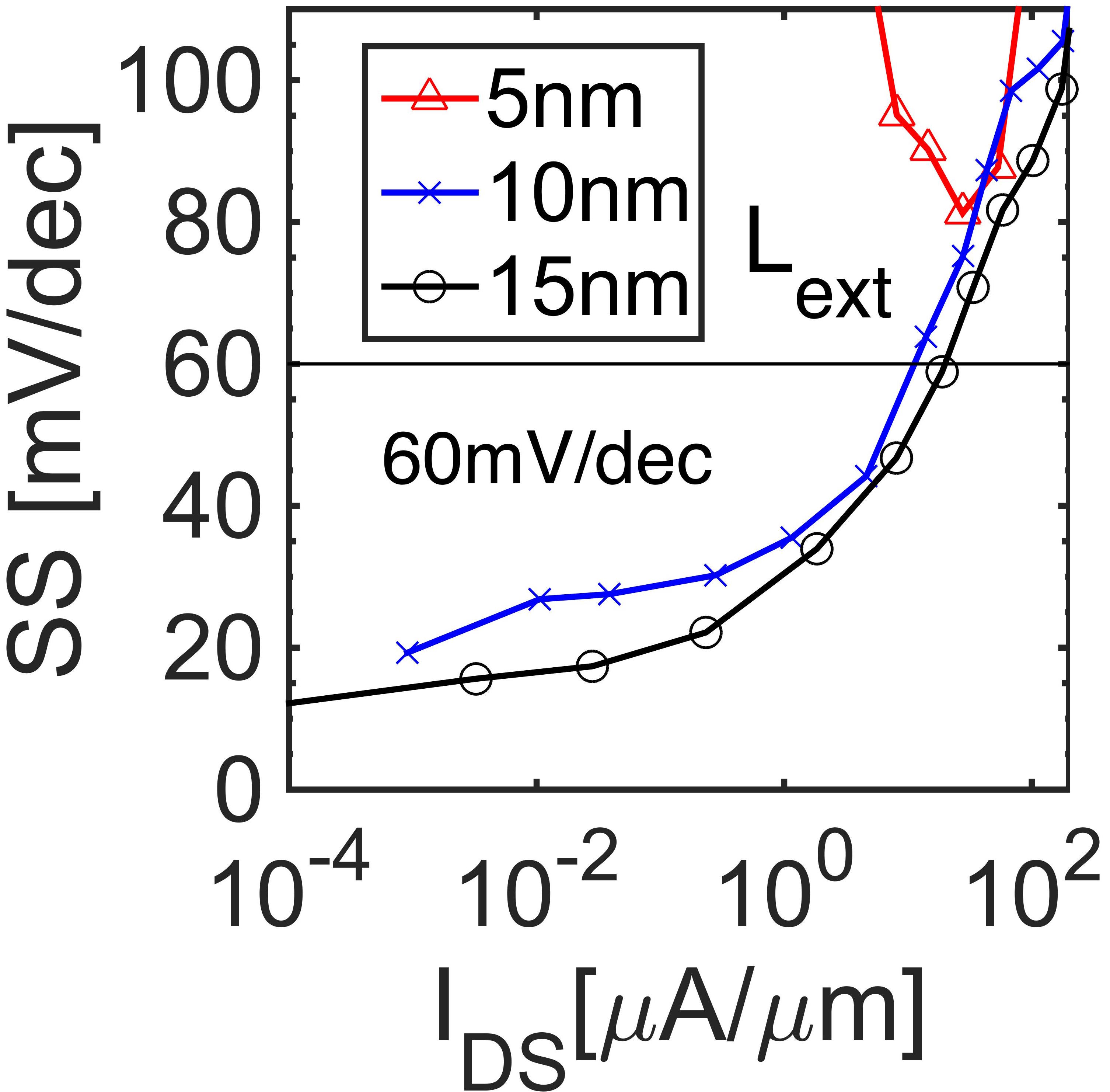}%
\label{3b}}
\vfil
\subfloat[]{\includegraphics[width=0.2\textwidth]{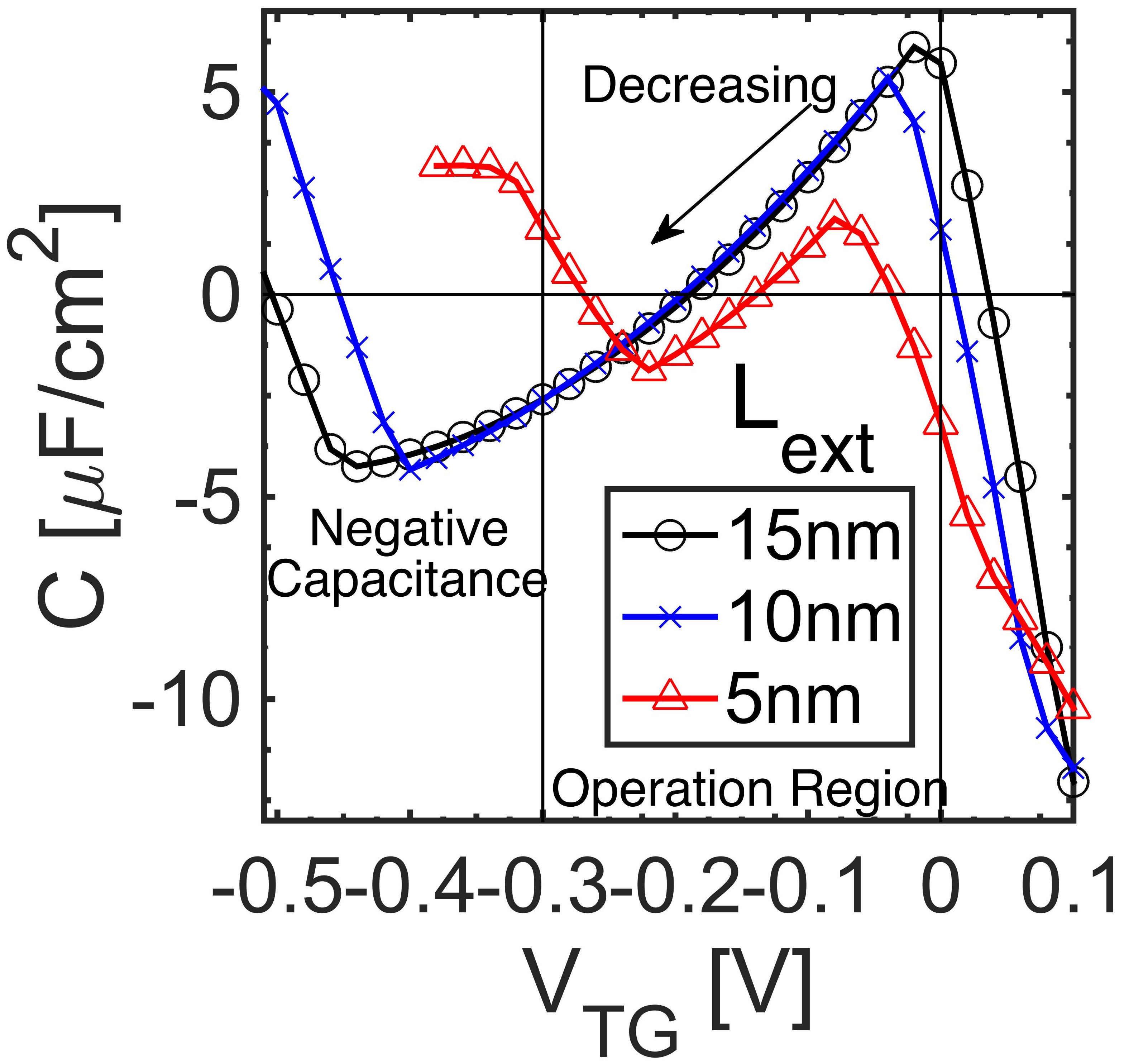}%
\label{3c}}
\hfil
\subfloat[]{\includegraphics[width=0.19\textwidth]{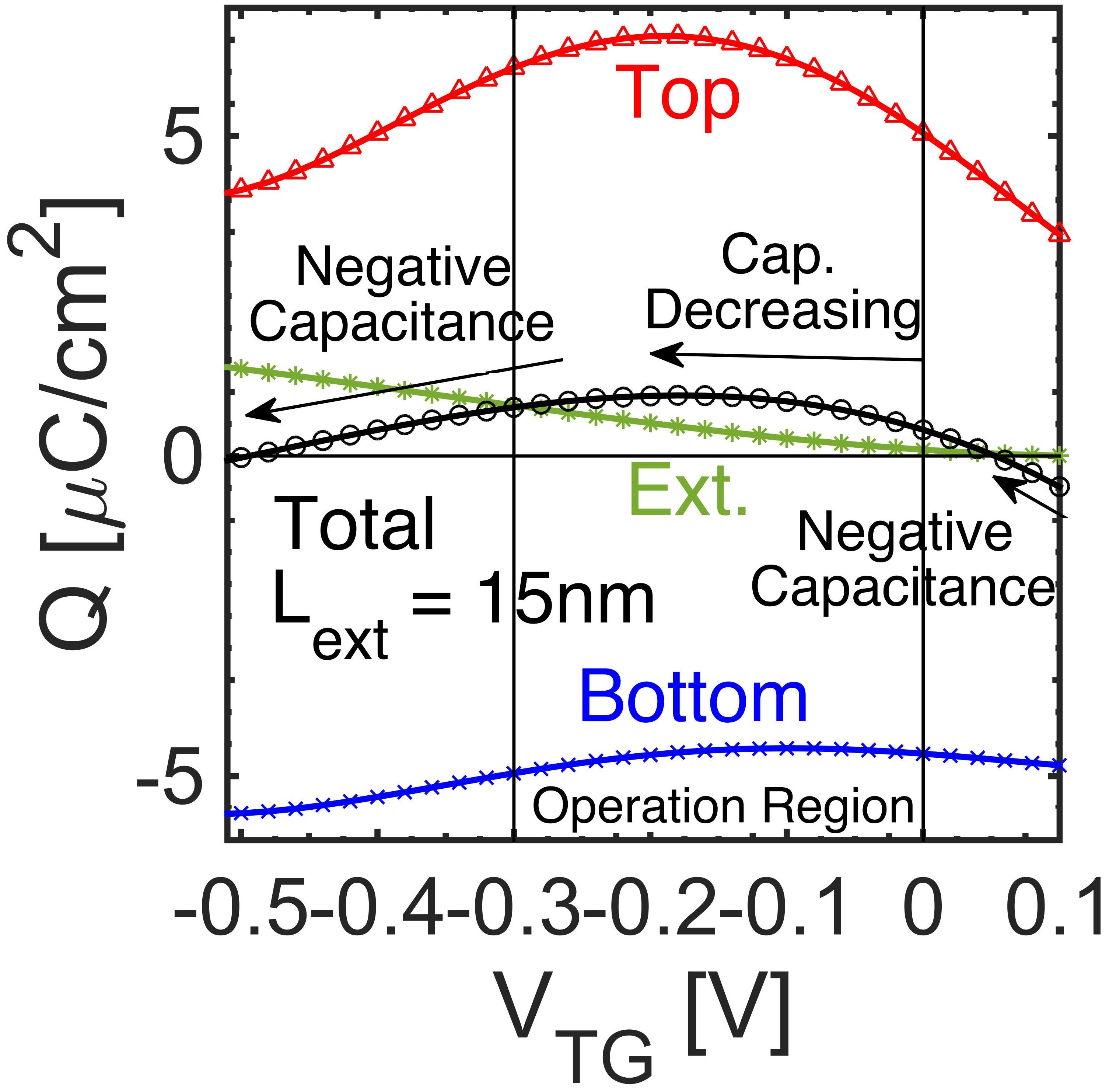}%
\label{3d}}
\caption{(a) The transfer characteristics ($I_d-V_g$); (b) The $SS-I_d$ plot and (c) The capacitance-voltage plot of a $MoS_2$-$WTe_2$ interlayer TFET with the extension length $L_{ext}$ ranging from 5 to 15nm. (d)The Charge-voltage plot of a $MoS_2$-$WTe_2$ interlayer TFET with $L_{ext} = 15nm$. Red, blue, green and black represent the integrated charge for top, bottom, extension layers and the whole channel respectively. }
\label{3}
\end{figure}

\begin{figure}
\centering
\subfloat[]{\includegraphics[width=0.22\textwidth]{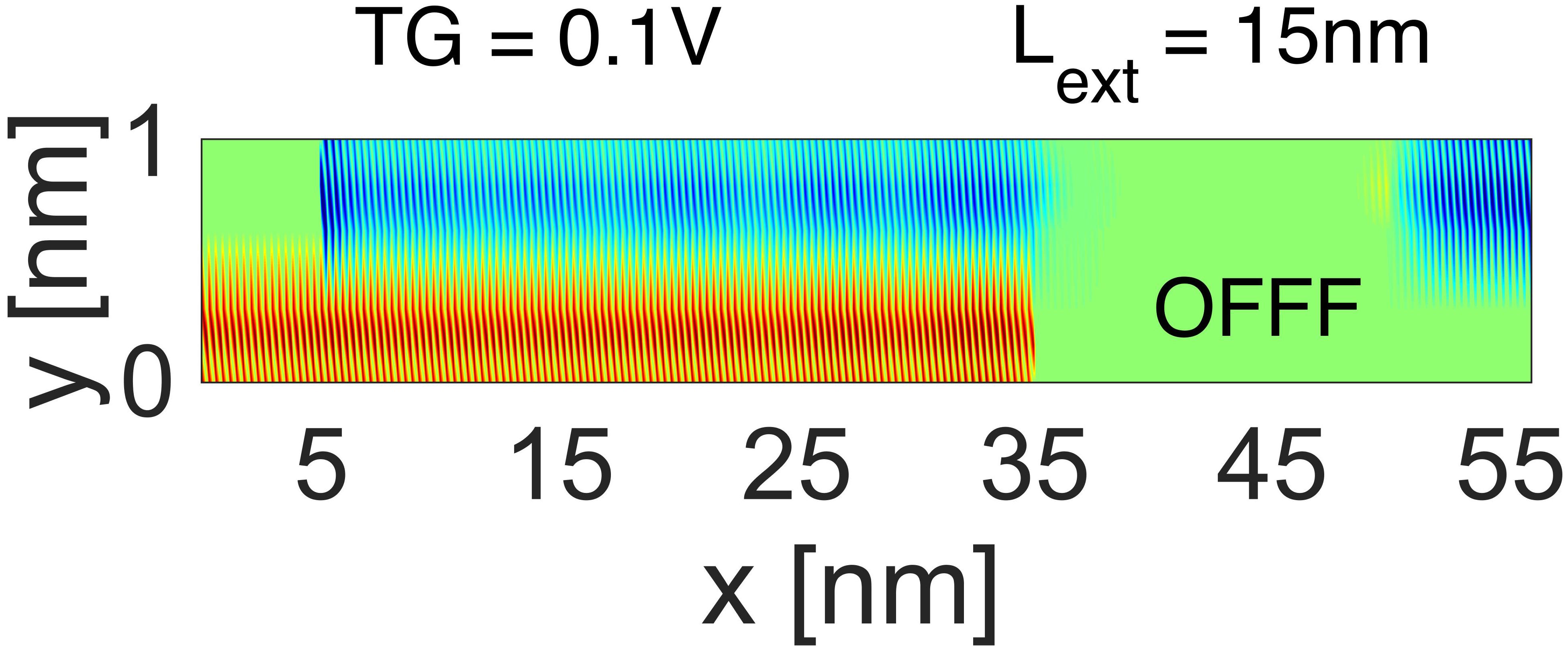}%
\label{4a}}
\hfil
\subfloat[]{\includegraphics[width=0.22\textwidth]{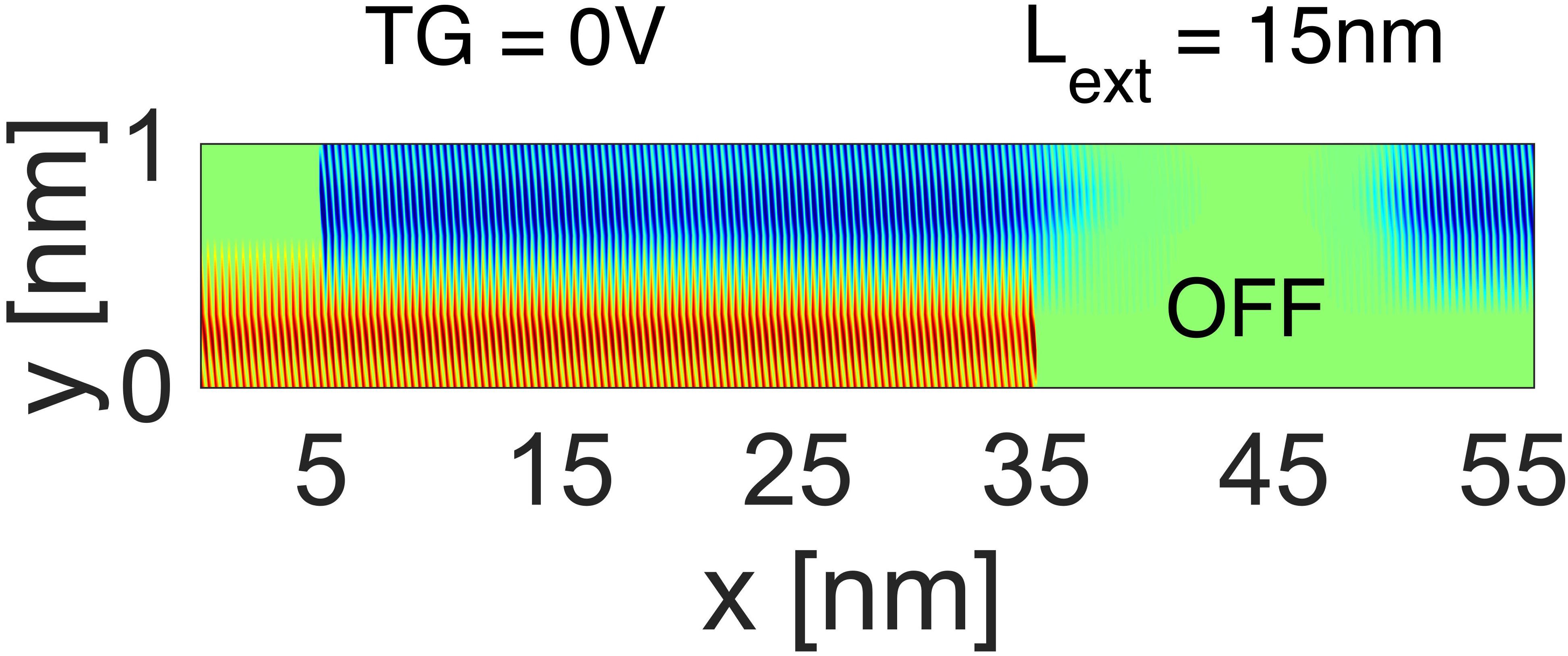}%
\label{4b}}
\vfil
\subfloat[]{\includegraphics[width=0.22\textwidth]{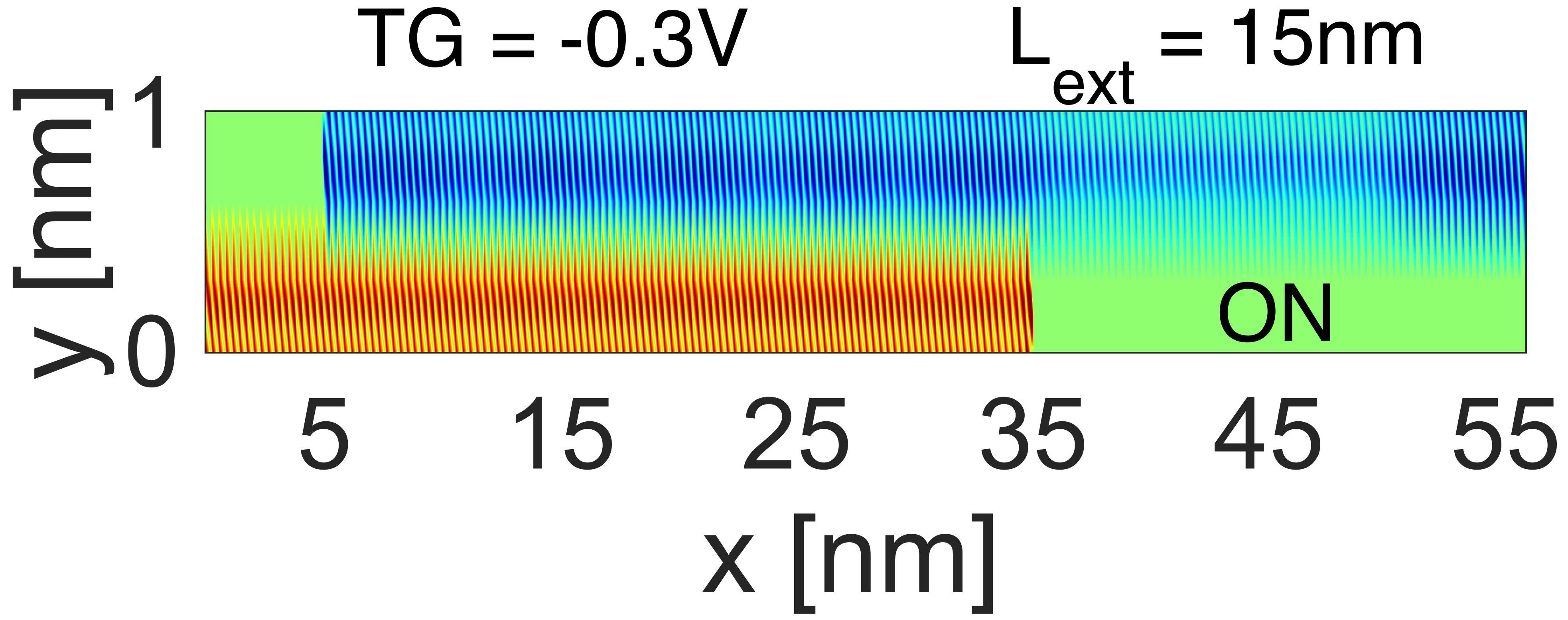}%
\label{4c}}
\hfil
\subfloat[]{\includegraphics[width=0.221\textwidth]{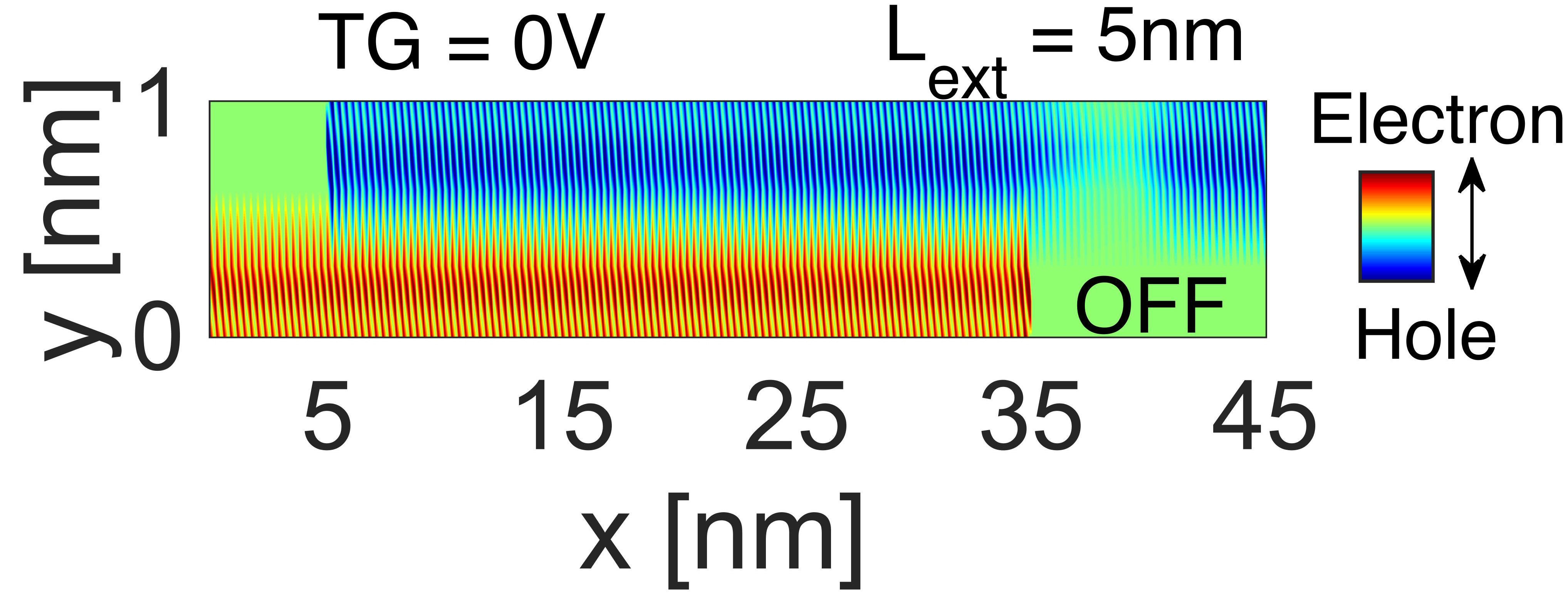}%
\label{4d}}
\caption{Position resolved electron/hole density for $MoS_2$-$WTe_2$ interlayer TFETs with $L_{ext} = 15nm$ at (a) OFFF state when $V_{TG} = 0.1V$ (outside device operation region); (b) OFF state when $V_{TG} = 0.0V$; and (c) ON state when $V_{TG} = -0.3V$; (d) ($L_{ext} = 5nm$) OFF state when $V_{TG} = 0.0V$. Red and blue represents the electron and hole density respectively.}
\label{4}
\end{figure}

 
 The band edges of a $MoS_2$-$WTe_2$ v-TFET and the corresponding energy resolved current at ON and OFF states respectively are plotted in Fig.~\ref{2}. The color contour of carrier density is plotted with respect to the energy and position. At both ON and OFF state, the carrier densities in the overlap region remain high. The carrier density in the extension region is changed from low to high when the device is turned ON. 
   
   The commonly believed switching mechanism of this vertical TFET is due to shifting of the relative band edges of the two materials as shown in Fig.~\ref{1b}, which bands of the top layer bands have shifted due to a bias of $V_{TG}$. However, as shown in the plot of position resolved diagram, the bands in the overlap region barely changed from OFF to ON. From this carrier density plot and also the 2D charge plot in Fig.~\ref{4}, there are always excessive charges populating the overlap region, which screens out the gate field. Whereas in the extension region, where only one layer of $WTe_2$ is present, the carriers can be tuned just like a conventional MOSFET.  The lack of conducting states in this extension region enables the device to turn off. In the meanwhile, the bands in the overlap region filters out the hot electrons. This is the reason of steep slope. This switching mechanism can be viewed as an energy filtering type \cite{smith2011broken}.\\
 
  Fig.~\ref{3a} shows the OFF current is degraded by reducing $L_{ext}$ from 15nm to 5nm. Besides the low OFF current, the extension region also helps the device to achieve small SS as shown in Fig.~\ref{3b}. Long extension regions are more efficient in blocking the source-drain current. As also shown in Fig.~\ref{4b}, for a device with $L_{ext} = 15nm$, the OFF current is totally blocked by the sufficiently long extension region. Whereas for a device with $L_{ext}=5nm$ as shown in Fig.~\ref{4d}, the current is leaking due to tunneling into the extension region with the same voltage applied. This concludes that a sufficiently long extension region is critical for v-TFETs to achieve both a low OFF current and a steep slope. 

Fig.~\ref{3c} shows that a negative capacitance for v-TFET is obtained under certain bias regimes. The Capacitance-Voltage (C-V) trends for all the devices with 5-15nm extension length are similar. The negative capacitance is the result of the simultaneous presence of charge of opposite polarity inside the channel as shown in the 2D charge plot of Fig.~\ref{4}(a)-(c). From OFFF state to ON state, the hole density in the bottom layer is barely modulated, whereas the electron density in the top layer increases as shown in Fig.~\ref{3d}. The initial increase in the top layer electron density neutralizes the hole density in the bottom layer, resulting in a decrease of the total net charge and a negative capacitance. \\
\begin{figure}[!t]
\centering
{\includegraphics[width=0.28\textwidth]{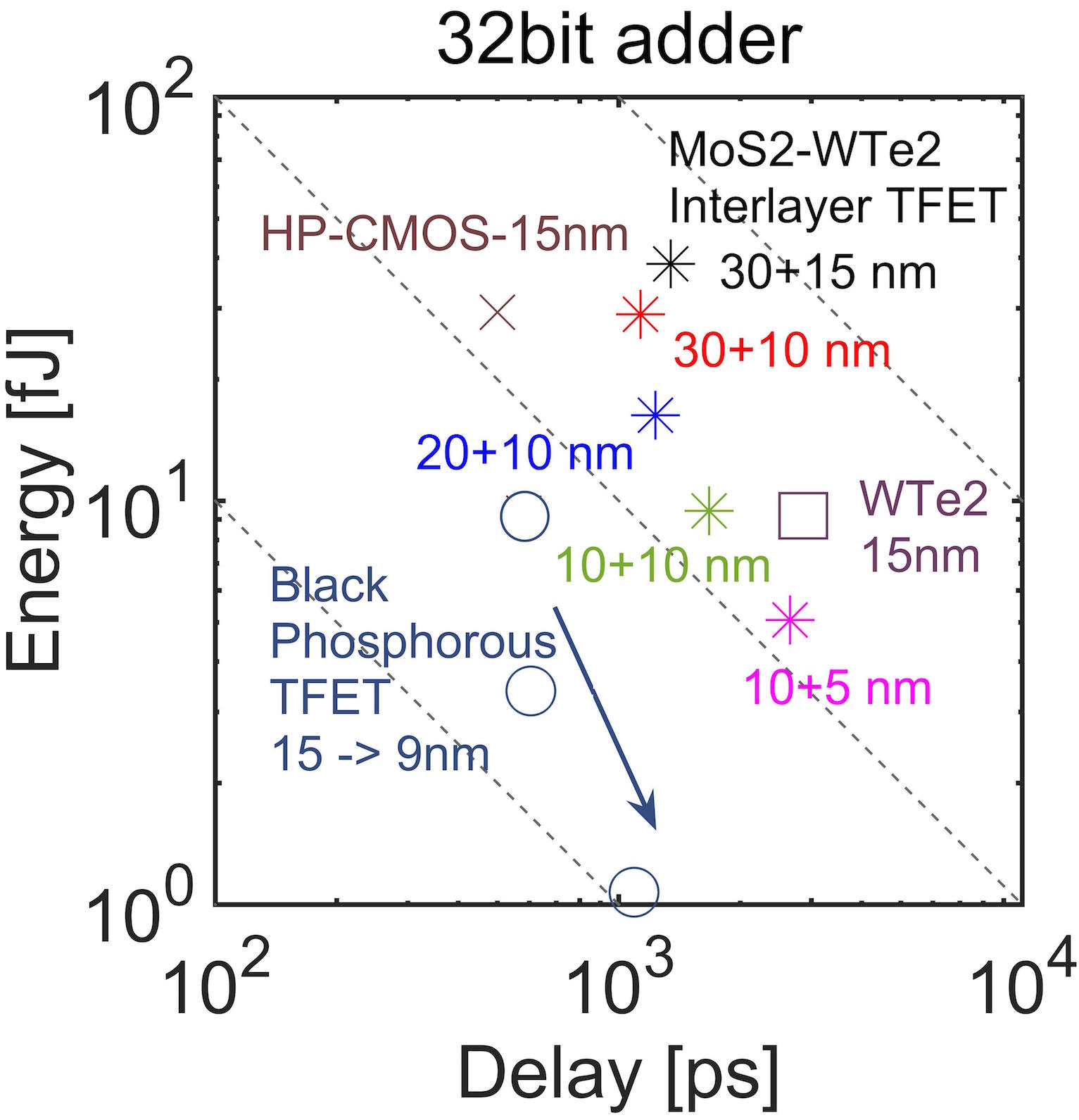}%
\label{60}}
\caption{Energy-Delay product of $MoS_2$-$WTe_2$ (star) interlayer TFETs with device channel scaled down from 55nm to 15nm in comparison to BP-TFETs (circle) \cite{ameen2015few, chen2017thickness} with their $L_{ch}$ scaled from 15nm to 9nm, a 15nm $WTe_2$ TFET (square) \cite{21} and high performance CMOS (cross). }
\label{6}
\end{figure}

 The $L_{ext}$ has to be kept longer than 10nm to maintain both the OFF and SS performance. If the $L_{overlap}$ is scaled, the ON current would be scaled linearly with it. The energy-delay product of a 32-bit adder composed of $MoS_2$-$WTe_2$ interlayer TFETs scaled from $30+15nm$ to $5+10nm$ with operation voltage $V_{DD} = 0.3V$ is plotted and compared with BP-TFETs \cite{ameen2015few, chen2017thickness} of $L_{ch}$ scaled from 15nm to 9nm, a 15nm $WTe_2$ TFET \cite{21} and high performance CMOS in Fig.~\ref{6}. The NMOS and PMOS performances are assumed to be the same in this circuit. The 32-bit adder energy delay product is calculated by using BCB 3.0 \cite{18}, where the parasitic capacitances and interconnects are taken into account. The scaling of circuit parameters as a function of gate length is taken from ITRS 2011 roadmap.

Although the $MoS_2$-$WTe_2$ v-TFET stays charged even during the OFF state, the negative capacitance resulting from the top layer charge neutralizing the bottom layer charge compensates the total capacitance. As a result, in the energy-delay product comparison, $MoS_2$-$WTe_2$ interlayer TFETs show a bit improvement from $WTe_2$ and HP-CMOS. The overall energy-delay performance of interlayer TFETs is also close to BP-TFET. This suggests v-TFETs as a promising candidate for future low power applications.\\

\section{Conclusion}
In conclusion, the switching mechanism and the prospects of scaling of $MoS_2$-$WTe_2$ v-TFETs are investigated. From the band diagram and the charge density plots, the switching mechanism is shown to arise from energy filtering. The transfer characteristics of the $MoS_2$-$WTe_2$ v-TFET further supports this. Due to this switching mechanism, an extension region being longer than 10nm is necessary to preserve both the SS and OFF state performance. This, however, stands in the way of device scaling. The presence of opposite charges simultaneously inside the channel results in a negative capacitance. This aids the energy-delay performance and makes v-TFETs one of the promising candidates for low power applications. \\

\bibliographystyle{IEEEtran.bst}

%
%
%

\bibliography{all.bib}

\end{document}